\documentclass[twocolumn,pra,amsmath,amssymb,unsortedaddress,superscriptaddress,showpacs]{revtex4}

\usepackage{latexsym,epsfig,graphicx}% Include figure files
\usepackage{dcolumn}% Align table columns on decimal point
\usepackage{bm}% bold math

\def\ket#1{\vert#1\rangle}
\def\bra#1{\langle#1\vert}

\begin{document}

\title{Structure and stability of Mott-insulator shells of bosons trapped in an optical lattice}

\affiliation{University of Illinois at Urbana-Champaign, Urbana,
IL 61801, USA} \affiliation{Wellesley College, Wellesley,
Massachusetts 02481, USA}

\author{B. DeMarco}
\affiliation{University of Illinois at Urbana-Champaign, Urbana,
IL 61801, USA}
\author{C. Lannert}
\affiliation{Wellesley College, Wellesley, Massachusetts 02481,
USA}
\author{S. Vishveshwara}
\author{T.-C. Wei}
\affiliation{University of Illinois at Urbana-Champaign, Urbana,
IL 61801, USA}

\date{\today}


\begin{abstract}
We consider the feasibility of creating a phase of neutral bosonic
atoms in which multiple Mott-insulating states coexist in a shell
structure and propose an experiment to spatially resolve such a
structure.  This spatially-inhomogeneous phase of bosons, arising
from the interplay between the confining potential and the
short-ranged repulsion, has been previously predicted. While the
Mott-insulator phase has been observed in an atomic gas, the
spatial structure of this phase in the presence of an
inhomogeneous potential has not yet been directly probed. In this
paper, we give a simple recipe for creating a structure with any
desired number of shells, and explore the stability of the
structure under typical experimental conditions. The stability
analysis gives some constraints on how successfully these states
can be employed for quantum information experiments. The
experimental probe we propose for observing this phase exploits
transitions between two species of bosons, induced by applying a
frequency-swept, oscillatory magnetic field.  We present the
expected experimental signatures of this probe, and show that they
reflect the underlying Mott configuration for large lattice
potential depth.
\end{abstract}

\maketitle

\section{Introduction}

Recent experimental results involving quantum degenerate atom
gases trapped in optical lattices have stimulated interest from
the perspectives of condensed matter and quantum information
science. Ultra-cold atoms confined in an optical lattice are
predicted to display a rich variety of quantum phases (see
\cite{Jaksch98,Kuklov04a}, for example). The ability to precisely
control the physical parameters of this system enables probing a
vast range of related fundamental physics including quantum phase
transitions, the behavior of collective excitations in the quantum
regime, and the physics of defects.  The properties of these
phases may permit new applications, such as neutral atom quantum
computing or quantum simulation \cite{Jane03}.

Neutral bosons trapped in optical lattices can exhibit complex
spatial configurations of co-existing superfluid and
Mott-insulator phases \cite{Jaksch98}. There have been detailed
studies of the superfluid to Mott-insulator phase transition in
the atomic system using techniques such as quantum Monte Carlo
simulation and mean-field theory (see
\cite{Oosten01,Kashurnikov02,Batrouni02,Wessel04} for examples).
In this paper we offer a practical and simple recipe for realizing
specific Mott-insulator configurations and we study the
limitations posed by realistic experimental conditions. Motivated
by proposals for using the Mott-insulator state to initialize a
fiducial state for neutral atom quantum computing
\cite{Rabl03,Brennen03a}, we focus on the tight-binding regime
where inter-site tunneling is suppressed. We formulate a
straightforward method, primarily using counting arguments, for
obtaining the constraints on lattice parameters that yield states
of definite spatially varying occupation number.

The effects of finite temperature, tunneling, and dissipation in a
periodic potential have been studied in the literature
\cite{LeggettRMP,Sethna,HanggiTalknerBorkovec90}. Atomic systems
are a unique tool for studying outstanding questions regarding
these phenomena. Furthermore, deviations from the zero-temperature
and tightly-bound ground state will have an important impact on
potential applications using atoms trapped in a lattice.  We
investigate processes which disturb the stability of the
Mott-insulator ground state and estimate their influence for
characteristic experimental conditions. Specifically, we initiate
an analysis of the emergence of superfluid order when there is a
finite but small degree of tunneling and we study
thermally-activated defects and their dynamics.  Our work is not
meant to be an exhaustive study of stability.  Instead, we take
the first steps toward addressing these considerations and point
to existing literature from the condensed matter community
appropriate to the atomic system.

To complement our theoretical work, we propose a set of
experiments using microwave spectroscopy to map out the spatial
structure of the bosonic states hosted in the optical lattice. The
proposed techniques can identify Mott-insulator states with any
site occupation number and resolve the spatial distribution of
sites with different occupation number. We view this method as
intermediate between less general techniques \cite{Rom04} and
high-resolution direct imaging \cite{Vala03} which will be
necessary for long-term development of neutral atom quantum
computing.

This paper focuses on experimental parameters similar to those
first used to observe the superfluid to Mott-insulator phase
transition in an atomic gas \cite{Greiner02a}. We consider using
$^{87}$Rb atoms confined in an optical lattice formed from three
pairs of intersecting laser beams with wavelength
$\lambda=850$~nm. The resulting lattice has cubic
three-dimensional symmetry with lattice spacing $a=\lambda/2$. The
effect of an external potential that changes over many lattice
sites, formed from magnetic or optical fields, is included.  The
lattice depth is $V_0=30 E_{\rm R}$, where the recoil energy
$E_{\rm R}$ is approximately $h\times3.7$~kHz ($h$ is Plank's
constant). The resulting tunneling strength between sites is of
order $w\approx h\times10$~Hz. The on-site interaction $U$ between
particles can be changed by changing $V_0$ or using a Feshbach
resonance.  For $V_0=30 E_{\rm R}$ and at zero magnetic bias
field, $U\approx h\times2.5$~kHz.

The outline of this paper is as follows. In Sec.~\ref{sec:Mott},
we present  a scheme for realizing Mott-insulator configurations
with specified occupation numbers. In Sec.~\ref{sec:stability}, we
address the stability of the Mott-insulator configuration, in
particular, against superfluid order and defect formation. In
Sec.~\ref{sec:experiment}, we describe experiments that exploit
microwave transitions between hyperfine states to probe the
Mott-insulator structure. We conclude with a summary of our
discussions in Sec.~\ref{sec:conclusion}.

\section{Co-existence of Mott-insulator phases} \label{sec:Mott}

In this section, we describe an optical lattice in a radially
symmetric geometry that permits the co-existence of concentric
Mott-insulating regions with differing occupation numbers. We
derive the constraints on lattice parameters necessary for this
co-existence. In contrast to previous work \cite{Jaksch98}, we
employ counting arguments to describe the spatial structure of the
Mott-insulator state, which allows us to consider a general
spatial geometry. Our starting point is the Bose-Hubbard
Hamiltonian \cite{FisherWeichmanGrinsteinFisher89,Jaksch98}
\begin{eqnarray}
\label{eqn:BH}
\hat{H} &=& \left\{ \sum_i \left[V(\textbf{r}_i)
- \mu \right] \hat{n}_i  + \frac{U}{2} \sum_i
\hat{n}_i \left( \hat{n}_i -1\right)\right\} \nonumber \\ & &-w \sum_{\langle i,j\rangle} \left( \hat{b}_i^{\dag} \hat{b}_j +
\hat{b}_j^{\dag} \hat{b}_i \right) \equiv \hat{H}_0 + \hat{W}
\end{eqnarray}
where $\mu$ is the chemical potential, $V(\textbf{r}_i)$ is the
value of the external confining potential at site $i$ of the
optical lattice (at the point $\textbf{r}_i$ in space),
$\hat{b}^{\dag}_i$ and $\hat{b}_i$ are the boson creation and
destruction operators at site $i$, $\hat{n}_i \equiv
\hat{b}_i^{\dag} \hat{b}_i$ is the number operator for bosons on
site $i$, $U$ is the on-site interaction energy between two
bosons, and $\hat{W}$ represents the tunneling that tends to
delocalize bosons, with $w$ being the tunneling matrix element for
bosons between nearest neighbor sites $\langle i,j\rangle$ on the
lattice.

In this section, we concentrate on the limit $w/U =0$ where
tunneling is negligible and $\hat{H} \rightarrow \hat{H}_0$. (For
small tunneling $w/U$, $\hat{W}$ can be treated as a perturbation;
we return to this in a later section.) The sites are decoupled in
this limit and the Hamiltonian, $\hat{H}_0$, is diagonalized by
the set of states with specific occupation number on each site
$\ket{n}_i={\big(b_i^\dagger\big)^{n_i}}\ket{0}_i/{\sqrt{n_i!}}$.
For a given $\ket{n}_i$, the energy is
$E_n=-\tilde{\mu}\,n+\frac{U}{2}(n^2-n)$, where $\tilde{\mu}$ is
an ``effective'' local chemical potential $\tilde{\mu}\equiv
\mu-V(r)$. The occupation number $n$ that minimizes the energy is
determined by ${\partial E_n}/{\partial
n}=0=-\tilde{\mu}-\frac{U}{2}(2n-1)$, which gives
 $n=\frac{\tilde{\mu}}{U}+\frac{1}{2}$.
Since $n$ can take on only integer values, the minimization
condition becomes
\begin{equation}
n-1<\frac{\tilde{\mu}}{U}< n, \ \ \mbox{$E_n$ is minimum}.
\end{equation}
The state with $n$ bosons, which we refer to as the $n$
Mott-insulator state, is an incompressible Mott-insulator phase.

\begin{figure}[t!]
\includegraphics[scale=1]{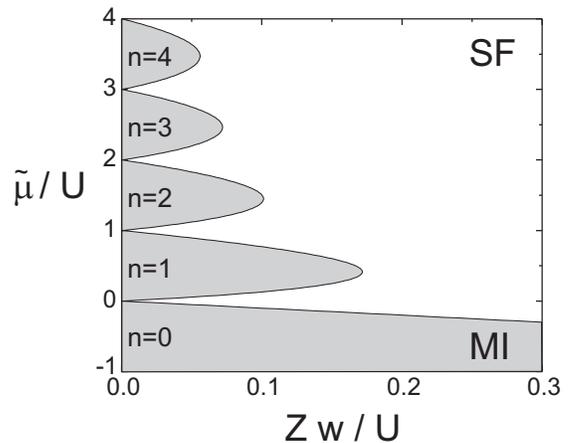}
\caption{The mean-field phase diagram for the Bose-Hubbard
Hamiltonian at first perturbative order in $w/U$. Here,
$\tilde{\mu} \equiv \mu-V(r)$, as described in the text, and $Z$
is the coordination number of the lattice ($Z=6$ for the
three-dimensional cubic lattice under consideration here). The
shaded regions are Mott insulating with the indicated number of
bosons per site, $n$. The white regions are superfluid. Notice
that the superfluid regions at integer values of $\tilde{\mu}$
extend all the way to zero tunneling ($w/U=0$). While the exact
location of the boundaries between superfluid and Mott regions at
finite tunneling is different at higher orders in perturbation
theory, the qualitative features of the phase diagram persist
\cite{Sachdev}.} \label{fig:SFMott}
\end{figure}

With finite but small tunneling ($w/U\neq0$), the regions where
$\tilde{\mu}/U$ is nearly an integer become superfluid, leading to
the well-known phase diagram shown in Fig.~\ref{fig:SFMott}. At
finite temperature but negligible tunneling, all Mott-insulator
phases are replaced by normal fluid phases of bosons which display
thermally activated
hopping~\cite{FisherWeichmanGrinsteinFisher89}. At suitably low
temperatures  $kT \ll \Delta E$, where $\Delta E$ is the
excitation energy for adding or removing a particle, the system is
most likely to be found in its ground state, which is the
situation we consider in this section; we discuss the effects of
finite temperature and tunneling in subsequent sections.

\begin{figure}[t]
\epsfxsize=3in \centerline{\epsffile{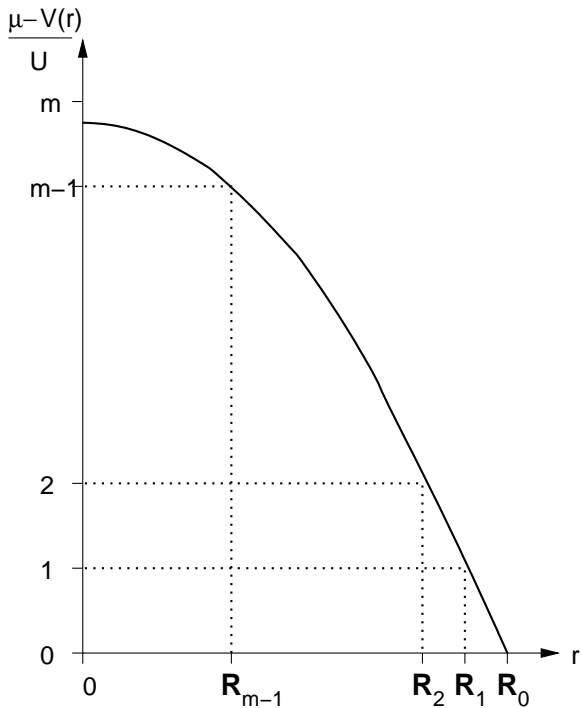}}
\caption{Illustration of the bounding values of $r$ for an
spherically-symmetric confining potential centered at $r=0$ which
supports shells of Mott-insulator phases.} \label{fig:Vr}
\end{figure}

For a spatially varying external potential $V(\textbf{r})$ , and
therefore spatially varying effective chemical potential
$\tilde{\mu}$, the system can have co-existing regions of
different Mott-insulator states~\cite{Jaksch98,GreinerThesis}. If
the external potential is spherically symmetric and of the form
$V(r) = \alpha r^{\xi}$, then $\tilde{\mu}$ is the largest at
$r=0$ and decreases for increasing $r$. Note that the arguments in
this section can be extended for more general potentials lacking
spherical symmetry. In the $w/U=0$ limit, the site occupation
number $m$ at the center is determined by the condition
$m-1<{\mu}/U<m$, where the chemical potential $\mu$ is determined
by the total number of particles $N$. The number of bosons per
site changes by one at each radius where $[\mu-V(\textbf{r})]/U$
is an integer, leading to a shell structure. The radii for the
boundaries, as illustrated in Fig.\ref{fig:Vr}, are given by $
R_n=\left([\mu-n U]/{\alpha}\right)^{1/\xi}$ ($n=0,1,\dots, m-1$),
where the $n$-th Mott-insulator state lies between $R_n$ and
$R_{n-1}$. This shell structure is established by the competition
between interaction and potential energy. The edge of the occupied
lattice is defined by the outermost boundary of the Mott-insulator
states where the occupation drops to zero
\begin{equation}
\label{eqn:r0} R_0=\left(\frac{\mu}{\alpha}\right)^{1/\xi}.
\end{equation}
All other radii can be expressed in terms of $R_0$:
\begin{equation}
R_n=R_0[1-n/(\mu/U)]^{1/\xi}. \label{eqn:rn}
\end{equation}

For a given interaction energy $U$ and total number of bosons $N$,
co-existence of a total of $m$ Mott shells will occur only for a
specific range of the parameter $\alpha$. We find that range as
follows: the chemical potential $\mu$ is determined via the global
constraint
\begin{equation}
N = \sum_i n_i(\mu).
\end{equation}
If the spacing between sites in the optical lattice, $a$, is small
compared with $R_0$, we can approximate each Mott region by a
three-dimensional continuous spherical shell with inner radius
$R_n$, outer radius $R_{n-1}$ and uniform density $n/a^3$. In this
continuum limit,
\begin{eqnarray}
N &=& \frac{1}{a^3} \int n(r) d^3r \\ &=& \frac{4\pi}{3a^3} \left(
R_0^3+R_1^3+R_2^3+ \ldots R_{m-1}^3 \right)
\\ &=&\frac{4\pi}{3a^3}R_0^3\sum_{n=0}^{m-1}\left(\frac{\mu-n
U}{\alpha}\right)^{3/\xi}.
\end{eqnarray}
With some rearrangement, we arrive at the condition
\begin{equation}
\label{eqn:xmu}
x\equiv\frac{3Na^3}{4\pi}\left(\frac{\alpha}{U}\right)^{3/\xi}
=\sum_{n=0}^{m-1}(\mu/U-n)^{3/\xi}.
\end{equation}
Because $m-1<\mu/U<m$, we can obtain from Eq.~(\ref{eqn:xmu}) the
upper and lower bounds on $x(\mu/U)$ necessary to realize a shell
structure with any desired maximum site occupancy $m$. For
example, for $\xi=2$ (a harmonic external potential),
\begin{eqnarray}
1&<&\frac{3Na^3}{4\pi}\left(\frac{\alpha}{U}\right)^{3/2}<3.82, \
\ m=2,\\
3.83&<&\frac{3Na^3}{4\pi}\left(\frac{\alpha}{U}\right)^{3/2}<9.03,
\ \ m=3.
\end{eqnarray}

As an example, we consider the experimental parameters detailed in
the introduction and $N = 10^6$ atoms.  To support three
Mott-insulator shells in a harmonic trap $V(r)=\alpha r^2$ the
curvature of the trap needs to satisfy the constraint
$h\times8.8$~Hz$/\mu \mbox{m}^2<\alpha<h\times15.6$~Hz$/\mu
\mbox{m}^2$. For a typical value in this range, for instance,
$\mu/U=2.5$ (and hence $\alpha=h\times12.1$~Hz$/\mu \mbox{m}^2$),
Eq.~(\ref{eqn:r0}) and Eq.~(\ref{eqn:rn}) show that the size of
the entire Mott-insulator structure is given by $R_0=22.8~\mu$m,
and that the radii separating the three Mott shells are
$R_1=17.6~\mu$m and $R_2=10.2~\mu $m.

In this section, we have shown how the spatial structure of the
Mott-insulator ground state in the absence of tunneling depends on
a limited number of experimental parameters.  Based on counting
arguments, we have derived a series of constraints on these
parameters for realizing a shell structure with a specific core
occupancy.

\section{Stability of the Mott-insulator phases}
\label{sec:stability}

The stability of the Mott structure is relevant to experiments
probing the physics described in section \ref{sec:Mott}.  Of
particular practical concern is the elimination of number
fluctuations when using the Mott-insulator state as a fiducial
state for neutral atom quantum information experiments.  Several
proposals suggest creating the Mott-insulator state followed by a
purification scheme to prepare a well-defined state with one atom
per lattice site \cite{Rabl03,Brennen03a}. These schemes cannot
eliminate sites that are initially vacant and suffer inaccuracy
when number fluctuations become large.

Deviations from the nested Mott-insulator shell structure will be
driven by finite tunneling between lattice sites and finite
temperature. The dynamic evolution of the system is a result of
the complex interplay between the Bose-Hubbard Hamiltonian,
dissipation, and finite temperature. While a complete study of
stability is beyond the scope of this paper, we summarize the
effects of tunneling and finite temperature and provide estimates
relevant to a typical experiment.

We assume that the lattice parameters described in the previous
section determine the initial conditions for the bosons and that
the lattice potential is free of imperfections.  In the following
sections, we first discuss global deviations from the
Mott-insulator structure coming from the tendency to form
superfluid regions.  We show that the effects of superfluidity can
be made negligible in the optical lattice. We then discuss the
local fluctuations caused by ``particle" and ``hole" (p-h) defects
in the Mott-insulator structure. We analyze the emergence and
dynamics of p-h defects at finite temperature and tunneling.

\subsection{Superfluidity} \label{superfluidity}

In principle, even the smallest amount of tunneling can alter the
Mott-insulator phase. At finite tunneling, the natural source of
instability is the formation of superfluid regions.
 As shown in the phase diagram of Fig.~\ref{fig:SFMott},
the system is particularly susceptible to forming regions of
superfluid at the boundaries between Mott states of different
occupation number. For a specfic set of values for the tunneling
$w$, interaction energy $U$ and external potential curvature
$\alpha$, we can estimate the size of the superfluid regions
formed between Mott phases. The Mott states remain relatively
robust against delocalization if the expected thickness of the
superfluid shells is less than a lattice spacing.

To estimate the size of the superfluid region at $T=0$, we invoke
a mean-field treatment that decouples lattice sites even in the
presence of tunneling~\cite{Sachdev}. To summarize the treatment,
the effective Hamiltonian takes the form
\begin{equation}
H_{{\rm MF}}  = \sum_i\Big[-\tilde{\mu}_i \hat{n}_i+  \frac{U}{2}
\hat{n}_i ( \hat{n}_i -1)- \Psi_B \hat{b}_i^{\dag}-\Psi^*_B
\,\hat{b}_i \Big].
\label{eqn:MF}
\end{equation}
Here $\Psi_B=Z w\langle b \rangle$ is related to the superfluid
order parameter and is obtained self-consistently, with $Z$ being
the co-ordination number (the number of nearest neighbors per
site). Terms involving $\Psi_B$ and $\Psi^*_B$ can be treated as a
perturbation to the Mott-insulator states $\ket{n}$. To second
order in $\Psi_B$, the perturbed eigenstates at a single site are
\begin{eqnarray}
\ket{\Psi_n} & = & \ket{n} + \Psi_B \alpha_1^+\sqrt{n+1}\ket{n+1}
+ \Psi_B^* \alpha_1^-\sqrt{n}\ket{n-1} \nonumber \\
 & & + \Psi_B^2\alpha_2^+\ket{n+2} +
 (\Psi_B^*)^2\alpha_2^-\ket{n-2},
\label{eqn:pertwvfn}
\end{eqnarray}
where $\alpha^{\pm}_{1/2}$ are functions of $n$, $\mu$ and $U$.
The energies are shifted to
\begin{equation}
\tilde{E}_n= E_n+\chi_n\big|\Psi_B\big|^2+{\cal
O}\Big(\big|\Psi_B\big|^4\Big), \label{eqn:pertenergy}
\end{equation}
where $E_n$ is the energy of the unperturbed Mott state $\ket{n}$,
and $\chi_n=\frac{n+1}{Un-\mu}+\frac{n}{\mu-U(n-1)}$.
Self-consistently minimizing $\tilde{E}_{n_0}- Z w\langle b
\rangle \langle b^{\dag} \rangle-\langle b\rangle \Psi_B^*-
\langle b^\dagger\rangle \Psi_B =
E_{n_0}+\rho\big|\Psi_B\big|^2+{\cal
O}\Big(\big|\Psi_B\big|^4\Big)$ determines $\Psi_B$. Here, $n_0$
is the Mott ground state occupation number appropriate for
$\tilde{\mu}$.  The superfluid boundaries are determined by the
sign of $\rho(\mu,w,U)$;  $\langle b \rangle$ has a non-zero
expectation value where $\rho(\mu,w,U)$ is negative. We have to
first non-vanishing order in $Zw/U$,  $\rho=\chi_{n_0}(\mu/U)[1-Zw
\chi_{n_0}(\mu/U)]$ with
\begin{equation}
\chi_{n_0}=\frac{n_{n_0}(\mu/U)+1}{U
n_0(\mu/U)-\mu}+\frac{n_0(\mu/U)}{\mu-U [n_0(\mu/U)-1]}.
\end{equation}
The superfluid boundary satisfies the condition
$1-Zw\chi_{n_0}=0$. For fixed tunneling, this determines the
boundary values of the effective chemical potential along each
Mott-insulator lobe in Fig.~\ref{fig:SFMott}. For $Zw/U\ll 1$, we
have for the effective chemical potential at the upper
($\tilde{\mu}^+_{n_0}$) and lower ($\tilde{\mu}^-_{n_0}$)
boundaries of the $n_0$ Mott-insulator lobe
\begin{equation}
\frac{\tilde{\mu}_{n_0}^+}{U}\approx n_0(1-Zw/U), \ \
\frac{\tilde{\mu}_{n_0}^-}{U}\approx (n_0-1)(1+Zw/U).
\label{eqn:mubound}
\end{equation}
Eq.~(\ref{eqn:mubound}) can be used to determine the radii
$r_{n_0-1}^+$ and $r_{n_0}^-$ between which superfluidity will
exist in the continuum limit. If the distance between the radii
where superfluidity can exist is smaller than a lattice spacing
($\Delta r=r_{n_0-1}^+-r_{n_0}^-  < a$), then number fluctuations
due to superfluid tendencies are expected to be negligible.

As an example, we again consider the experimental situation
detailed in the introduction with a harmonic external potential
and $\mu/U=2.5$. The thickness of the superfluid between the
$n_0=1$ and $n_0=2$ Mott-insulator states is $\Delta
r_{1-2}\approx0.82\sqrt{{U}/{\alpha}}\,{Zw/U}$. Using the value
$\alpha\approx h\times12.1$~Hz$/\mu \mbox{m}^2$ from the previous
section, we obtain  $\Delta r_{1-2}\approx0.07~\mu$m and $\Delta
r_{2-3}\approx0.24~\mu $m, which is smaller than the lattice
spacing  $a\approx0.43~\mu $m. Therefore, for this specific case,
no superfluid regions will exist. Eq.~(\ref{eqn:mubound}) could
also be used to determine the experimental parameters leading to
the existence of superfluid shells, which may be of fundamental
interest in and of themselves.

\subsection{Defects}

While global instabilities coming from a tendency for the bosons
to condense can be made negligible, local instabilities can still
permeate the Mott-insulator structure in the form of defects.  In
this section, we discuss two types of local excitations
---  isolated defects
where a site has an extra particle or one less (hole) compared
with the Mott-insulator ground state, and excitations where a
particle has hopped to a neighboring site and leaves behind a
hole. The former, which we call ``site-defects" can arise from
random removal of atoms from the lattice, for example by heating
from spontaneous emission driven by the lattice light or by
collisions with room-temperature residual gas molecules in the
vacuum system. We refer to the latter as ``particle-hole" (p-h)
excitations and these are natural perturbations to the
Mott-insulator ground state. Site-defects can also arise when
components of p-h excitations (due to finite temperature or
imperfect adiabatic transfer) that were present during initial
loading of the Bose-Einstein condensate into the lattice become
widely separated.

In the following discussion, we first consider the energy gap
associated with these excitations; the energy gap plays a key role
in the statics and dynamics of defects. We then study the static
perturbative effects of p-h excitations on the Mott-insulator
state in the presence of finite tunneling. Next, we consider the
dynamics of defects within the framework of a toy two-site model.
We comment on defect dynamics over the entire system and on the
effects of dissipation. Finally, we discuss the role of finite
temperature.

{\it \bf Energy Gap}  In the Mott-insulator ground state at zero
tunneling, the functional $E(n)=-\tilde{\mu}n+\frac{U}{2}n(n-1)$
is minimized at each site. The energy gap associated with single
site defects is $\Delta E^+=E_{n_0+1}-E_{n_0}=-\tilde{\mu}+U n_0$
for the addition of an extra particle and $\Delta
E^-=E_{n_0-1}-E_{n_0}=\tilde{\mu}-U (n_0-1)$ for the removal of a
particle (addition of a hole), where $n_0$ is the ground-state
occupation number. Since $n_0-1<\frac{\tilde{\mu}}{U}<n_0$,
$\Delta E^{\pm}$ is positive and of order $U$ in the bulk of the
system. Close to Mott boundaries, though, the energy gap becomes
arbitrarily small as $\frac{\tilde{\mu}}{U}$ approaches either
$n_0$ or $n_0-1$. Because of the underlying lattice and
inhomogeneous confining potential, however, the energy cost for a
site defect rarely vanishes: there is always a change in potential
energy $\Delta V$ over a lattice spacing and it is unlikely for a
boundary to exactly coincide with a lattice site. For the
experimental parameters detailed in the introduction,
 $V_i=(h\times 12.1~{\rm Hz}/{\mu {\rm m}^2})
r_i^2$, we have $\Delta V=h\times181$~Hz ($h\times104$~Hz) at the
boundary  $R_1=17.6~\mu {\rm m}$ ($R_2=10.2~\mu {\rm m}$).

The energy cost for a p-h excitation in the bulk is
\begin{equation}
\label{eqn:hopB}
E_i^{{\rm hop}}(n_0)\equiv \Delta E_i^++\Delta E_{i+1}^-=\Delta V_{i}+U,
\end{equation}
where $i$ is a site index, $\Delta V_i = V_{i}-V_{i+1}$, and $V_i$
is the on-site potential energy from the confining potential. For
a particle hopping across a boundary between a Mott-insulator
$n_0$ phase on site $i$ and $n_0-1$ phase on site $i+1$, the
energy cost is
\begin{eqnarray}
E_i^{{\rm hop}}(n_0\downarrow) & \equiv & \Delta E_i^-(n_0)+\Delta E_{i+1}^+(n_0-1)=\Delta V, \nonumber \\
E_i^{{\rm hop}}(n_0\uparrow) & \equiv & \Delta E_i^+(n_0)+\Delta E_{i+1}^-(n_0-1)\nonumber \\
&=&2U-\Delta V,
\end{eqnarray}
The symbol $\downarrow$ denotes hopping from an $n_0$ to an
$n_0-1$ Mott-insulator state and $\uparrow$ denotes the reverse
process. The most favorable p-h excitation therefore involves a
particle hopping across a Mott-insulator boundary from higher to
lower occupation.

{\it \bf Finite Tunneling - Statics:}

At small tunneling ($w\ll U$), corrections to the $w=0$
Mott-insulator ground state $\ket{\psi_0}$ can be written in terms
of p-h excitations provided that $w\ll E_i^{\rm hop}$. As found in
the previous discussion, since the smallest $E_i^{\rm hop}$ is
across Mott boundaries with \mbox{$E_i^{{\rm
hop}}(n_0\downarrow)$}$=\Delta V$, the change in potential energy
must be much larger than the tunneling energy ($\Delta V\gg w$),
which is achieved by the parameters under consideration, namely,
$w\approx h \times 10$~Hz and $\Delta V = h \times 181 $~Hz ($h
\times 104$~Hz) across the boundary at $R_1=17.6~\mu$m
($R_2=10.2~\mu$m). This inequality is related to the condition
found in Sec.~\ref{superfluidity} for the putative superfluid
regions to occupy less than a lattice site.

To first order in $w/E^{\rm hop}$, the perturbative groundstate
$\ket{\psi}$ of Eqn.(\ref{eqn:BH}) is given by
\begin{equation}
\ket{\psi}=c\left(\ket{\psi_0}+\sum_{{\rm exc}}\ket{{\rm
exc}}\frac{\bra{{\rm exc}}\hat{W}\ket{\psi_0}}{\Delta E^{\rm
exc}}\right),
\end{equation}
where $c$ is a normalization constant. The excitations and
corresponding activation energies are $\ket{\rm exc}$ and $\Delta
E^{\rm exc}$, respectively, and the sum is over all p-h
excitations. Comparing $\ket{\psi}$ and $\ket{\psi_0}$, we can
estimate the deviation of the many-body state or the single site
occupancy from the zero-tunneling Mott-insulator ground state.

As an example, consider the case of a finite-sized uniform lattice
with no external potential and occupancy $n_0$ at each site in the
absence of tunneling. To lowest order, tunneling perturbs
 the Mott-insulator ground state $\ket{\psi_0}=\ket{n_0,n_0,...n_0}$ to
\begin{eqnarray}
\label{eqn:Psi}
&&\ket{\psi}=c\left( \frac{}{} \ket{n_0,\cdots,n_0}+\right. \\
&& \frac{w\sqrt{n_0(n_0\!+\!1)}}{U}
\sum_{\langle i,j\rangle}\ket{n_0,\cdots,(n_0\!\pm\!1)_i,(n_0\!\mp\!1)_j\cdots n_0}
%+\ket{n_0,\cdots,(n_0\!+\!1)_i,(n_0\!-\!1)_j\cdots n_0}
\left. \frac{}{} \right). \nonumber
\end{eqnarray}
Deviations of this many-body state from $|\psi_0 \rangle$ can be
quantified by the zero-temperature
fidelity~\cite{NielsenChuang00}, defined as
\begin{equation}
f_0=|\langle{\psi}\ket{\psi_0}|^2.
\label{eqn:fzeroT}
\end{equation}
$1-f_0$ is essentially the probability that the system is not in
the zero-temperature Mott-insulator ground state. For the
wavefunction in Eqn.(\ref{eqn:Psi}), the fidelity takes the form
\begin{equation}
f_0\approx\Big[1+\frac{w^2 n_0(n_0\!+\!1) Z\, N_s}{U^2}\Big]^{-1},
\label{eqn:fpsi}
\end{equation}
where $N_s$ is the total number of occupied sites. A high fidelity
requires the possibly severe requirement $\sqrt{N_s Z}n_0 w/U\ll
1$. For experiments involving local probes, however, the more
relevant quantity is the defect probability,
 $P_{i,w}^{{\rm def}}$, at a given site $i$. The probability for a site to be
in the state $\ket{n_0\pm1}$ can be obtained from
Eq.~(\ref{eqn:Psi}):
\begin{eqnarray}
\label{eqn:PDef}
 P_{i,w}^{{\rm def}}\approx \Big[\frac{Z \,w^2 n_0(n_0\!+\!1)}{U^2}\Big]/\Big[1+\frac{w^2 n_0(n_0\!+\!1) Z\,N_s}{U^2}\Big].
\end{eqnarray}
Low defect probability locally, then requires $\sqrt{N_sZ} w
n_0/U\ll 1$, which is satisfied for the experimental parameters
considered in the introduction.

Modifying these arguments for the Mott-insulator shells in an
inhomogeneous potential is straightforward. The number of p-h
excitations varies spatially with the energy gap for excitations.
The fidelity will therefore contain terms of order $\Delta V$ from
sites at boundaries and contributions of order $U$ from sites in
the Mott-insulator bulk. The condition for low defect probability
for sites in the bulk is the same as for the uniform case. The
probability of a defect, $\tilde{P}_{i,w}^{\rm{def}}$, for a
boundary site becomes of order $w\bar{n}_0/\Delta V \ll 1$, where
$\bar{n}_0$ is the average number of bosons per site near the
boundary.

{\it \bf Finite Tunneling - Dynamics:} To analyze the evolution of
a configuration of defects in time and the emergence of p-h
excitations driven by quantum dynamics, we use a two-site version
of the Bose-Hubbard Hamiltonian:
\begin{equation}
H=-w(b_1^\dagger b_2+b_2^\dagger
b_1)+\frac{U}{2}\sum_{i=1,2}n_i(n_i-1)+\Delta V n_2.
\label{eqn:BH2}
\end{equation}
Here, $i=1,2$ are site indices and $\Delta V$ is the change in
on-site potential between sites $1$ and $2$.

As the simplest case, we consider the restricted Hilbert space of
just one particle present, i.e., states $\ket{1,0}$ and
$\ket{0,1}$ in the $\ket{n_1,n_2}$ basis. This situation could
correspond site-defects composed of either a hole in the $n=1$
bulk or a particle in the $n=0$ bulk, or to the boundary between
the $n=1$ and $n=0$ states. For $\Delta V =0$, corresponding to
site-defects at fixed radius in the external potential, the
quantum state tunnels back and forth between $\ket{1,0}$ and
$\ket{0,1}$ at a rate $w/h$. For $\Delta V \ne 0$ and the particle
initially in state $\ket{1,0}$, the probability for the particle
to remain in state $\ket{1,0}$ is
\begin{equation}
P^{(1)}_{1\rightarrow 1}(t)=1-\frac{w/\Delta V}{\left[1+(2w/\Delta
V)^2\right]}(1-\cos\alpha_1 t), \label{eq:prob}
\end{equation}
where $\alpha_1=\sqrt{\Delta V^2+4w^2}/\hbar$. The probability to remain
in $\ket{1,0}$ is depleted at most by $w/\Delta V$ for $\Delta
V\gg w$, which is true for the experimental conditions described
in the introduction.

Interactions come into play when there are at least two particles
and the Hilbert space is confined to $n_1+n_2=2$. To focus on the
effect of interactions, we set $\Delta V=0$. The state $\ket{1,1}$
corresponds to a state in the $n=1$ in the Mott-insulator bulk,
while $\ket{0,2}$ and $\ket{2,0}$ correspond to p-h excitations.
Only the states $\ket{11}$ and $\ket{\psi_{2+}}\equiv
(\ket{20}+\ket{02})/\sqrt{2}$ are
 coupled via the hopping term. The probability
$P^{(2)}_{1\rightarrow1}(t)$ for particles to remain in the state
$\ket{11}$ is identical to $P^{(1)}_{1\rightarrow 1}(t)$ from
Eq.~(\ref{eq:prob}) with $U$ replacing $\Delta V$ and double the
value of $w$. The probability for $\ket{11}$ to decay into p-h
excitations is at most of order $w/U$, which is $4\times10^{-3}$
for the experimental parameters discussed in the introduction. The
same probability and rate hold for the decay of any p-h
excitations already present.

We expect the most pronounced long-range dynamics to come from
site-defects tunneling along sites of equipotential. In fact,
since the kinetic energy of a site-defect is reduced by tunneling,
site-defects will tend to delocalize, corresponding to
superfluid-like behavior. As the simplest possibility, we can
model a particle in the $n=0$ Mott bulk or a hole in the $n=1$
Mott-insulator bulk moving along an equipotential surface by
setting the potential energy $\Delta V$ and the interaction energy
$U$ to zero in the Bose-Hubbard Hamiltonian. In this purely
tight-binding limit, the amplitude for a particle starting at site
labelled by an integer vector $\mathbf{r}$ to propagate to a site
labelled by $\mathbf{s}$ is given by

\begin{eqnarray}
\bra{\mathbf{s}}e^{-iHt}\ket{\mathbf{r}}&=&\int \frac{d^dk}{(2\pi)^d}
e^{-i \mathbf{k}\cdot(\mathbf{r}-\mathbf{s}) + i\, 2w t\sum_{i=1}^d\cos k_i}\\
&=& e^{- \pi/2\sum_{i}\Delta r_i}\prod_{i=1}^d J_{\Delta r_i} (-2wt)
\end{eqnarray}
where $d$ is the effective spatial dimension, $\mathbf{k}$ is the
momentum, $\Delta \mathbf{r}\equiv (\mathbf{r}-\mathbf{s})$, and
$J_\nu(x)$ is the Bessel function of the first kind. The maximum
probability amplitude is transferred from $\mathbf{r}$ to
$\mathbf{s}$ on a time-scale
$t\approx{\hbar|\mathbf{r}-\mathbf{s}|}/(2w)$, suggesting that
site-defects can exhibit ballistic motion along the equipotential
surfaces.

We have so far neglected the effects of dissipation. One source of
dissipation is spontaneous emission driven by the lattice light
which will occur at relatively low (1-10~Hz) rates. We expect the
major source of dissipation at any given site to come from the
interactive coupling of the site to the entire system. The
calculation of such a dissipative term for optical lattices has
not been performed to our knowledge. Tunneling of site-defects
between nearest neighbors in the presence of dissipation can be
treated phenomenologically by the well-studied ``spin-boson''
Hamiltonian, which describes the physics of a particle in a double
well system coupled to an environment \cite{LeggettRMP}. We have
yet to analyze the effects of dissipation on defects in optical
lattices, and at present, we refer the reader to the exhaustive
spin-boson literature \cite{LeggettRMP}. In the analogous
situation of defects in solids, dissipation arising from coupling
to phonons is believed to be ``superohmic'', leading to weak
damping of the oscillations of the equivalent two-level system
\cite{LeggettRMP,Sethna}.

In the absence of dissipation, there is no complete tunneling
between ground and excited states. However, we would expect
dissipation to relax an excited metastable state to the ground
state over a long time scale $\tau_{{\rm
dis}}=[\gamma(\frac{w}{\Delta E})^2]^{-1}$, where $\gamma$ is the
dissipative rate and $\Delta E$ is the energy of the excited
state. Dissipation and decoherence of the site-defect wavefunction
can also result in a cross-over to classical behavior. In this
case, we expect site-defects to propagate diffusively rather than
ballistically.

{\it \bf Finite Temperature :}  The stability of the
Mott-insulator shell structure requires temperatures sufficiently
low for thermal excitations to be insignificant ($k_BT \ll U,
\Delta V$). The state at finite temperature is captured by the
statistical density matrix
\begin{equation}
\rho=\sum_{n} e^{-\frac{E_n}{k_BT}}\ket{E_n}\bra{E_n},
\end{equation}
where $k_B$ is Boltzmann's constant, $T$ is the temperature, and
$E_n$ and $\ket{E_n}$ are excitation energies and corresponding
 states, respectively. The finite temperature generalization of the
 fidelity given in Eq.~(\ref{eqn:fzeroT}) becomes
\begin{equation}
f=\langle{\psi_0}|\rho\ket{\psi_0}/{\rm Tr(\rho)},
\label{f-finiteT}
\end{equation}
where $\ket{\psi_0}$ is again the Mott-insulator ground state. At
low temperature, p-h excitations occur with Boltzmann weight and
the effect of tunneling can be accounted for perturbatively. The
finite temperature fidelity in Eq.~(\ref{f-finiteT}) for
deviations from the uniform Mott-insulator ground state is given
by
\begin{equation}
f(T)\approx f_0 \frac{1}{1+ Z N_s e^{-\frac{U}{k_BT}}},
\end{equation}
where $f_0$ is the fidelity at zero temperature.  Likewise, the
probability of finding a defect at a site $i$, given by
Eq.~(\ref{eqn:PDef}), becomes
\begin{equation}
P_{i,w}^{\rm def}(T)\approx [P_{i,w}^{\rm def}(0)+ Z
e^{-\frac{U}{k_BT}}]/[1+ZN_s e^{-\frac{U}{k_BT}}].
\end{equation}
Compared to zero temperature, the defect probability at finite
temperature becomes enhanced by a Boltzmann term appropriate for
p-h excitations. The defect probability is largest across a
boundary:
\begin{equation}
\tilde{P}_{i,w}^{\rm{def}}(T)\approx
\frac{\tilde{P}_{i,w}^{\rm{def}}(0) + e^{-\frac{\Delta
V}{k_BT}}}{1+ N_b e^{-\frac{\Delta V}{k_BT}}+ZN_s
e^{-\frac{U}{k_BT}}},
\end{equation}
where $N_b$ is the number of pairs of sites across the boundary,
$\Delta V$ is the potential difference, and
$\tilde{P}_{i,w}^{\rm{def}}(0)$ is the corresponding
zero-temperature defect probability.

\begin{figure}[t]
\includegraphics[scale=1]{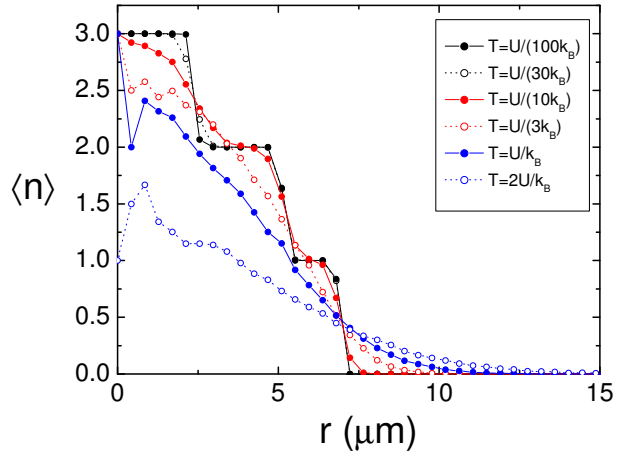}
\caption{Spatial profile of occupation number for six different
temperatures (see legend).  A Metropolis algorithm is used to
calculate the distribution of roughly $3 \times 10^4$ atoms among
lattice sites with a harmonic external potential characterized by
$\alpha=h\times113$~Hz$/\mu m^2$ and the experimental parameters
detailed in the introduction. The average number of atoms per site
$\langle n\rangle$ is shown at radii $r$ which are multiples of
the lattice spacing. The radius is measured from the minimum of
the external potential.} \label{fig:profile}
\end{figure}

\begin{figure*}[t]
\includegraphics[scale=1.3]{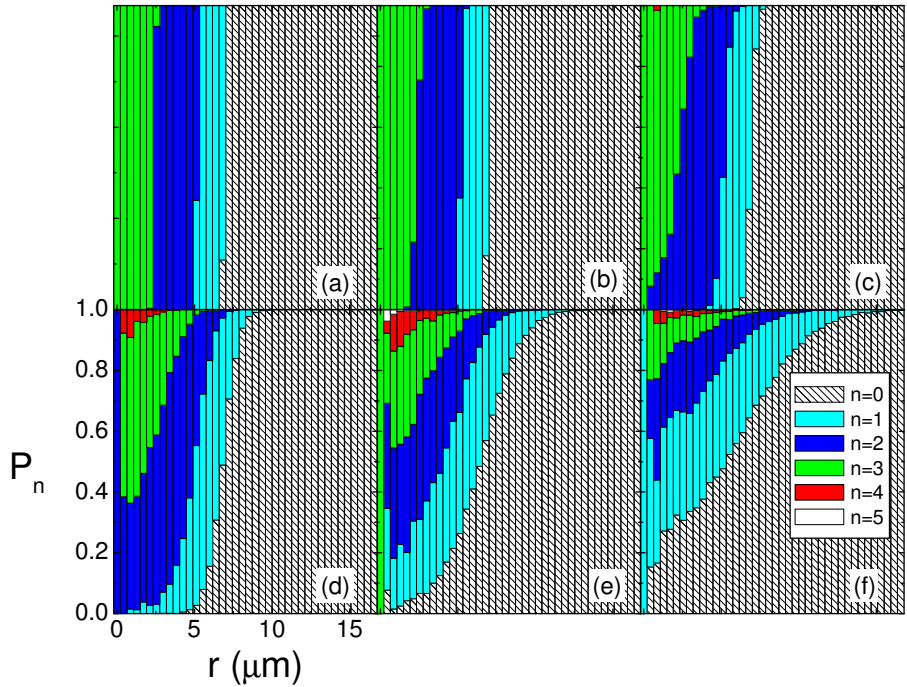}
\caption{Probability distribution of occupation number for six
different temperatures: (a) $U/(100k_B)$, (b) $U/(30k_B$), (c)
$U/(10k_B$), (d) $U/(3k_B)$, (e) $U/k_B$, and (f) $2U/k_B$. The
probability $P_n$ to find a site occupied by $n$ atoms is shown
for radii $r$ which are multiples of the lattice spacing. The data
are extracted from the Metropolis algorithm used in
Fig.~\ref{fig:profile} and are displayed as a stacked column plot
(see legend).} \label{fig:pops}
\end{figure*}

We have simulated the equilibrium configuration for a system of
$N\approx 3 \times 10^4$ atoms using a Metropolis algorithm, in
contrast to a slave-boson model employed in \cite{dickersheid03}.
Fig.~\ref{fig:profile} shows the spatial profile of occupation
number from this simulation for various temperatures. The
simulation indicates that the temperature must be smaller than
$U/(10k_B)$ for thermal fluctuations of particle number to be
substantially suppressed.  A plot of the probability distribution
for different site occupancies is shown in Fig.~\ref{fig:pops}.
Well-defined spatial regions where most sites are occupied by the
same number of atoms disappear for temperatures higher than
$U/(10k_B)$. The data in Fig.~\ref{fig:pops}(c) also show how
thermally activated hopping is most likely at boundaries between
the Mott-insulator states closest to the minimum of the external
potential.

The time scale for thermally activated hopping can be probed in
experiments and will play a role in schemes where a zero entropy
state is initialized for neutral atom quantum computing
\cite{Vala03}.  At the crudest level of approximation, the
tunneling rates due to quantum and thermal effects can be added to
obtain the net rate, particularly if the time scales are widely
separated~\cite{HanggiTalknerBorkovec90}. At zero temperature we
found that dynamic fluctuations of p-h excitations and site-defect
propagation across potential barriers are largely suppressed due
to interactions.  However, at finite temperature, these processes
can be thermally activated. To describe hopping between sites, one
can invoke a phenomenological double well model and appeal to the
exhaustive knowledge of reaction-rate theories for hopping between
metastable states~\cite{HanggiTalknerBorkovec90}.

\section{Microwave Spectroscopy}\label{sec:experiment}

In this section we describe a technique for probing the spatial
structure and energy spectrum of the Mott-insulator state
discussed in section \ref{sec:Mott}. The method, an extension of
work in \cite{Rabl03}, is compatible with any lattice geometry and
is capable of resolving the effects of finite temperature. In
contrast to the previous sections of this paper where only one
atomic internal state was considered, multiple hyperfine states
are employed in our spectroscopic technique. Transitions between
hyperfine states are driven using a microwave-frequency magnetic
field. Information on the interaction energy $U$ and the density
profile of the gas is obtained by manipulating the resonant
transition frequency between hyperfine states. Dependence on site
occupancy is achieved by changing the interaction energy $U$ using
a Feshbach resonance, while a spatially inhomogeneous,
state-dependent optical potential provides spatial resolution.

\subsection{Interaction with an oscillating magnetic field}

We first address the atomic interaction with an oscillating
magnetic field in anticipation of explaining the detailed
experimental procedure in subsequent sections. In the presence of
an oscillating magnetic field, the exact Hamiltonian at one
lattice site in the absence of tunneling is
\begin{eqnarray}
\left\{\hat{H}_{ab}\right\}+\hat{H}_I&=&\left\{\frac{U_{bb}}{2}\hat{n}_b(\hat{n}_b-1)+\right.\nonumber\\
&&\left.\frac{U_{aa}}{2}\hat{n}_a(\hat{n}_a-1)+U_{ab}\hat{n}_a\hat{n}_b+\hbar\omega_{ab}\hat{n}_b\right\}\nonumber\\
&&-\sum_i\hat{\vec{\mu}}^{(i)}\cdot\vec{B} \label{demarcoeq1}
\end{eqnarray}
where $\hat{n}_a$ and $\hat{n}_b$ are the number operators for
atoms in hyperfine states $|a\rangle$ and $|b\rangle$, $U_{aa}$,
$U_{bb}$, and $U_{ab}$ are the interaction energy for two atoms in
states $|a\rangle$ or $|b\rangle$ or one atom in each state, the
sum runs over the atoms, and $\hat{\vec{\mu}}$ is the atomic
magnetic moment operator. The energy difference $\hbar\omega_{ab}$
between states $|a\rangle$ and $|b\rangle$ takes into account the
effect of any external optical and magnetic potentials, including
a static magnetic bias field that defines the $z$ direction. We
also assume that the lattice potential is identical for both
hyperfine states; specifically we consider a far off-resonance
lattice.  The applied microwave-frequency magnetic field
$\vec{B}=B_l\cos\left[\omega(t)t\right]\hat{e}_x$ may have a
time-dependent frequency $\omega$. The interaction with the
magnetic field (the last term in Eq.~(\ref{demarcoeq1})) can be
rewritten as
$\hat{H}_I=\hbar\Omega[e^{i\omega(t)t}+e^{-i\omega(t)t}]\sum_i(\hat{\sigma}_+^{(i)}+\hat{\sigma}_-^{(i)})$
where $\hbar\Omega=-\mu_{ab}B_l/4$, $\mu_{ab}$ is the magnetic
moment matrix element between the $|a\rangle$ and $|b\rangle$
states, and $\hat{\sigma}_\pm^{(i)}$ are the Pauli raising and
lowering operators for atom $i$.

Because the states $|a\rangle$ and $|b\rangle$ form an effective
spin-1/2 system, the eigenstates of $\hat{H}_{ab}$ are eigenstates
of total angular momentum $\hat{J}=\sum_{i=1}^{n}\hat{S}^{(i)}$,
where $\hat{S}^{(i)}$ is the effective spin operator for atom $i$.
The interaction with the oscillating magnetic field, which can be
rewritten as
$\hat{H}_I=\Omega[e^{i\omega(t)t}+e^{-i\omega(t)t}](\hat{J}_++\hat{J}_-)$,
causes a rotation of the total spin vector. Since we assume that
all of the atoms start in one hyperfine state, $\hat{H}_I$ couples
states with different $m_j$, but fixed total angular momentum
quantum number $j=n/2$ ($n=n_a+n_b$ and
$J^2|j,m_j\rangle=\hbar^2j(j+1)|j,m_j\rangle$). All states with
$j=n/2$ are properly symmetrized with respect to two-particle
exchange for bosons.

We take advantage of the Schwinger representation and write the
eigenstates $|j,m_j\rangle$, which are states of $n$ spin-1/2
particles, as $|n_a,n_b\rangle$.  In this representation,
$\hat{J}_+|n_a,n_b\rangle=\sqrt{n_a(n_b+1)}|n_a-1,n_b+1\rangle$
and
$\hat{J}_-|n_a,n_b\rangle=\sqrt{n_b(n_a+1)}|n_a+1,n_b-1\rangle$.

The interaction picture is convenient for calculating the effect
of the oscillating field.  Taking into account that the
interaction Hamiltonian can only couple states where $n_a$ changes
by one, that $n_a+n_b$ is constant, and by making the
rotating-wave approximation, we obtain
\begin{eqnarray}
\hat{H}_I'&=&\hbar\Omega\left\{\hat{J}_+e^{-i[\omega(t)-(E_{n_a-1,n_b+1}-E_{n_a,n_b})/\hbar]t}+\right.\nonumber\\
&&\left.\hat{J}_-e^{i[\omega(t)-(E_{n_a,n_b}-E_{n_a+1,n_b-1})/\hbar]t}\right\}
\label{demarcoeq2}
\end{eqnarray}
in the interaction picture.  In Eq.~\ref{demarcoeq2}, the energy
$E_{n_a,n_b}|n_a,n_b\rangle=\hat{H}_{ab}|n_a,n_b\rangle$  is
determined by the eigenvalues of $\hat{H}_{ab}$. The quantum state
of the atoms on a lattice site can be written in terms of the
time-independent eigenvectors of $\hat{H}_{ab}$ as
\begin{equation}
\Psi=\sum_{n_a=0}^nC_{n_a,n_b=n-n_a}|n_a,n_b=n-n_a\rangle
\end{equation}
where $n=n_a+n_b$.  In general, the Schr\"{o}dinger equation gives
rise to coupled equations for the $C_{n_a,n_b}$ that can be solved
numerically.

To illustrate the salient features of this interaction
Hamiltonian, we will ignore off-resonant excitation and discuss
the case $\omega(t)=E_{n_a-1,n_b+1}-E_{n_a,n_b}+\delta(t)$.  While
for $^{87}$Rb the interaction energy matrix elements are normally
hyperfine-state-independent to within a few percent, we consider
using a Feshbach resonance so that we can have $U_{ab}=U_{aa}$ and
$U_{bb}=\beta U_{aa}$. We discuss only the states $n=0,1,2,3$ even
though the proposed experimental techniques will work for any site
occupancy. For typical experimental conditions, three-body
recombination rates will limit the lifetime of the $n>2$ state to
a few ms. However, we include analysis of that state here in light
of a recent proposal \cite{Meystre2004} for reducing three-body
recombination rates by a factor of 100.

\begin{figure}[t]
\includegraphics[scale=0.55]{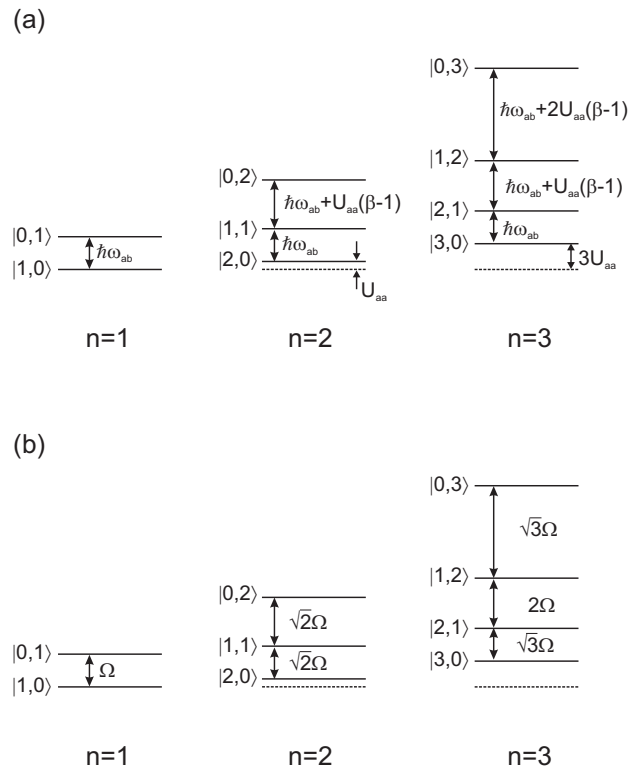}
\caption{The energy spectrum of $\hat{H}_{ab}$ for lattice sites
with $n=1,2,3$ (a) and the Rabi frequencies for transitions
between states that are coupled by the interaction Hamiltonian
(b). In (a), the energy difference between states and the shift of
the ground state are shown.} \label{demarcofig1}
\end{figure}

The energy spectrum of the eigenstates of $\hat{H}_{ab}$ for
states with different $n$ is shown in Fig.~\ref{demarcofig1}(a).
Solving the Schr\"{o}dinger equation for the atomic wavefunction
gives coupled equations of the form
\begin{eqnarray}
\dot{C}_{n_a,n_b}=-i\Omega_{n_a,n_b} C_{n_a+1,n_b-1}e^{-i\delta(t)}\\
\dot{C}_{n_a+1,n_b-1}=-i\Omega_{n_a,n_b} C_{n_a,n_b}e^{i\delta(t)}
\end{eqnarray}
where $\Omega_{n_a,n_b}=\sqrt{(n_a+1)n_b}\Omega$.  In the absence
of off-resonant excitation and for $\delta$ independent of time,
the problem reduces to oscillations between two states with Rabi
frequency $\Omega_{n_a,n_b}$ (shown in Fig.~\ref{demarcofig1}(b)).
There are two important features of the coupled problem that are
apparent in Fig.~\ref{demarcofig1}. The first is that there are no
degenerate energy splittings in any manifold for constant $n$.
However, the transition frequencies are degenerate when the
transition involves states with the same $n_b$ between manifolds
of different $n$. Also, the Rabi rates for transitions
$|0,n\rangle\leftrightarrow|1,n-1\rangle$ and
$|n,0\rangle\leftrightarrow|n-1,1\rangle$ are unique to each
manifold of constant $n$, regardless of the value of $\beta$
(i.e., $|1,1\rangle\leftrightarrow|0,2\rangle$ and
$|2,1\rangle\leftrightarrow|1,2\rangle$). In the next section, we
will describe two experiments that will take advantage of these
features.

\subsection{Experimental Proposals}

The two experimental techniques that we describe in this section
are a simple method for determining the ratios of sites with
different $n$, and a more developed technique for mapping out the
spatial structure of the insulator state. We consider experiments
using $^{87}$Rb atoms and the experimental parameters outlined in
the introduction. With the laser intensity required to reach
30~$E_R$ lattice depths, spontaneous emission rates driven by the
lattice light will be negligible on experimental timescales. The
limiting timescale for experiments will therefore be collisions
with residual gas molecules in the vacuum system or spontaneous
scattering from an additional optical potential.

\subsubsection{Rabi oscillations}

The first experiment that we propose uses Rabi oscillations as a
probe of the total population in the lattice of sites with any
$n$. A similar technique has been used in ion-trap experiments to
measure populations in different harmonic oscillator levels
\cite{Leibfried97a,Meekhof96a,BenKish2003}. The total number of
sites with $n$ atoms is identified by the Rabi oscillation
amplitude at the Rabi rate $\Omega_{n,0}$ for the
$|n,0\rangle\leftrightarrow|n-1,1\rangle$ transition. Because the
resonant frequency and Rabi oscillation rate for this transition
are independent of interaction energy, this method is insensitive
to the motional state of the atoms.

The gas is first prepared in the Mott-insulator state in the
lattice; harmonic confinement is provided by an inhomogeneous
magnetic field. The atoms in the lattice are initialized in the
state $|a\rangle$ (through optical pumping, for example). A
magnetic field at frequency $\omega_{ab}$ is applied to couple the
$|a\rangle$ and $|b\rangle$ states, and the population in
$|b\rangle$ is measured for different lengths of time of the
applied coupling (via resonance fluorescence, for example). The
data are fitted to a sum of oscillating functions with frequencies
that differ by a factor $\sqrt{k}$ where $k=1,2,3...$. The total
population in the lattice in states of different $n$ are then
directly connected to the amplitudes of each oscillating term in
the fit.

One complication with this scheme is that the inhomogeneous
magnetic field will introduce a spread in $\omega_{ab}$ across the
lattice, which will cause dephasing of the Rabi oscillations at
different sites.  To suppress shifts in $\omega_{ab}$ across the
lattice, the energy difference between states $|a\rangle$ and
$|b\rangle$ should be insensitive to magnetic field. This can be
accomplished using a first-order magnetic field insensitive
transition, for example between the states $|F=2,m_F=1\rangle$ and
$|F=1,m_F=-1\rangle$ near 3.24~G \cite{Hall98a}. For $10^6$ atoms,
to realize sites with $n=3$ will require at least a 7~mG spread
across the gas using either of these states. With a 3.24~G
magnetic bias field, there will be less than a $2\pi\times0.1$~Hz
shift in $\omega_{ab}$ across the occupied sites.

\subsubsection{Rapid Adiabatic Passage}

In this section, we explain an experimental technique to probe the
Mott-insulator spatial structure directly, where the occupation of
sites is identified using the interaction energy. Rapid adiabatic
passage is used to transfer atoms in state $|a\rangle$ to state
$|b\rangle$. The interaction energy is manipulated by using a
Feshbach resonance such that the resonant microwave frequency
shifts for transitions between $|a\rangle$ and $|b\rangle$ in a
manifold of states with constant $n$. The maximum frequency shift
dependence on $n$ permits unique identification of sites with
different $n$.

To change the interactions between atoms in state $|b\rangle$, we
choose $|b\rangle=|F=1,m_F=1\rangle$ and apply a uniform magnetic
field with magnitude close to the $^{87}$Rb Feshbach resonance at
1000.7~G \cite{Volz03,Marte02}.   The magnitude of the field is
adjusted so that $\beta=5$ ($U_{bb}=5U_{aa}$). Because this
particular Feshbach is exceptionally narrow, applying a uniform
magnetic field is necessary to avoid spatially-varying interaction
energy shifts. Therefore, rather than obtaining spatial
information by employing an inhomogeneous magnetic field, a
spatially inhomogeneous near-resonant optical potential is used to
shift $\omega_{ab}$. The near-resonant laser fields are engineered
to provide a harmonic external potential; this may be accomplished
using diffractive optics \cite{Curtis02,Sinclair04} to produce a
focused laser beam with a parabolic intensity profile in three
dimensions. To maximize the shift in $\omega_{ab}$ while
minimizing spontaneous emission rates, we choose
$|a\rangle$=$|F=2,m_F=2\rangle$ and tune the laser wavelength and
polarization so that the optical potential for the $|b\rangle$
state vanishes.  For different atoms, such as $^{85}$Rb, with
broader Feshbach resonances, the complication of employing an
optical potential may not be necessary.

\begin{figure}[t]
\includegraphics[scale=0.4]{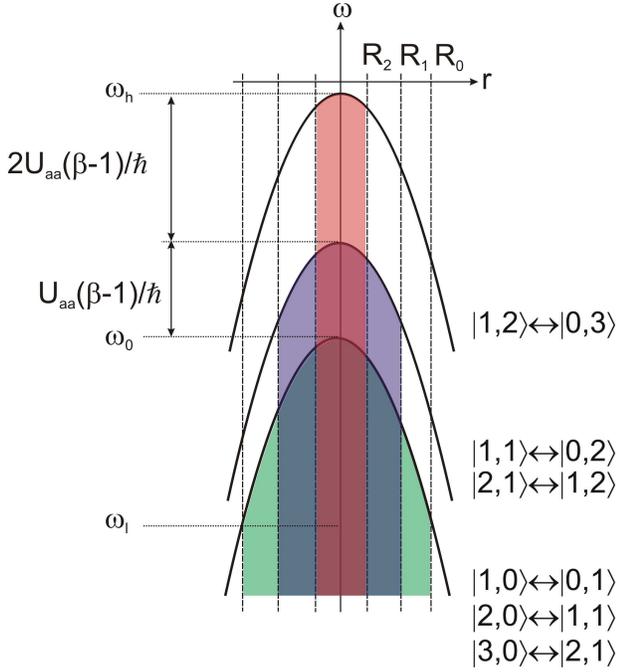}
\caption{The spectrum for transitions between $|a\rangle$ and
$|b\rangle$ for $n=1,2,3$. The transition frequency between states
with different $n_a$ and $n_b$ is shown by the thick line. The
dashed lines indicate the boundaries between sites with different
$n$ in the lattice, which occur at radii $R_0$, $R_1$, and $R_2$.
Sites with $n=1,2,3$ are indicated with green, blue, and red
coloring, respectively. The interaction energy shifts are
indicated, and the figure is drawn assuming
$\beta>0$.}\label{demarcofig2}
\end{figure}

The overall effect of the applied, parabolic optical potential is
to shift $\omega_{ab}=\omega_0-\gamma r^2$ parabolically in space.
The curvature $\gamma$ is related to the curvature of the
intensity profile of the near-resonant optical field,
$\omega_0\approx9.126$~GHz is determined by the atomic hyperfine
structure and the magnetic bias field, and $r$ is the distance
from the minimum of the parabolic optical potential. The spectrum
for transitions between states $|a\rangle$ and $|b\rangle$ for
sites with $n=1,2,3$ is shown in Fig.~\ref{demarcofig2}.  For our
experimental procedure, the curvature $\gamma$ is adjusted so that
the minimum interaction energy shift is at least equal to the
shift from the parabolic optical potential for the $n=2$ state at
the boundary with the $n=1$ state.  While this may be incompatible
with supporting the desired Mott-insulator structure, the applied
state-dependent potential can be switched quickly before the
spectroscopy described next is performed.

The experiment is performed by first preparing the atoms in the
$|a\rangle$ state.  Storing the atoms for long times in the
$|b\rangle$ state while the magnetic field is near the Feshbach
resonance is avoided to minimize effects due to inelastic
collisions.  A microwave frequency magnetic field is then swept
from $\omega_l$ to $\omega_h$ and the number of atoms transferred
to $|b\rangle$ is measured (using resonance fluorescence, for
example) as the frequency is changed.

Atoms are transferred to state $|b\rangle$ in stages.  As the
frequency is swept toward $\omega_h$, one atom at all occupied
sites is transferred to $|b\rangle$.  This process is interrupted
at $\omega_0$ by a gap where no atoms are transferred to
$|b\rangle$ as the frequency is changed.  The size of that gap in
frequency $U_{aa}(\beta-1)/\hbar-\gamma R_1^2$ is related to the
interaction energy, the size of the region where there are at
least two atoms per site, and the curvature $\gamma$ of the
optical potential. After this gap in atom transfer, one additional
atom per site is transferred to state $|b\rangle$ for all sites
that have at least two atoms per site.  Another gap is encountered
at $\omega_0+U_{aa}(\beta-1)/\hbar$ where no atoms are transferred
that is $2U_{aa}(\beta-1)/\hbar-\gamma R_2^2$ wide in frequency.
Finally, a third atom is transferred to $|b\rangle$ in each site
where $n=3$ as the frequency is swept to $\omega_h$. No atoms are
transferred to $|b\rangle$ as the frequency is increased past
$\omega_0+3U_{aa}(\beta-1)/\hbar$.  The interaction energy and the
distribution of sites in space width different occupancy $n$ can
be inferred from the number of atoms transferred to $|b\rangle$ as
the frequency is changed.

This sequence of transfer between hyperfine states separated by
gaps is generic to any Mott-insulator structure.  Transfer begins
at $\omega_{rf}=\omega_0-\gamma R_0^2$, which corresponds to the
boundary of the occupied sites at $R_0$. The width of each gap
where atoms are not transferred to $|b\rangle$ as the frequency is
changed is $kU_{aa}(\beta-1)/\hbar-\gamma R_k^2$, where
$k=1,2,3...$. The highest frequency where atoms is transferred is
determined by the occupancy $m$ at the core and is $\omega_0+m
U_{aa}(\beta-1)/\hbar$.

\begin{figure}[t]
\includegraphics[scale=0.90]{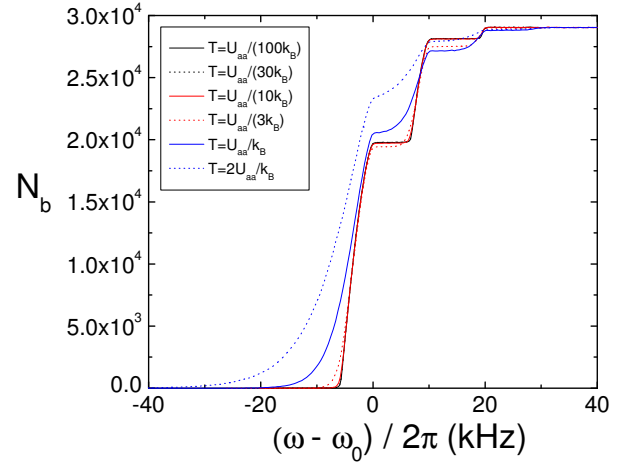}
\caption{Simulation of rapid adiabatic passage experiment for
roughly $3\times 10^4$ atoms at six different temperatures (see
the legend, where $k_b$ is Boltzmann's constant).  The number of
atoms $N_b$ in state $|b\rangle$ is shown as the frequency
$\omega_{rf}$ of an oscillating magnetic field is swept.
}\label{demarcofig3}
\end{figure}

The results of a simulation of this experimental technique is
shown in Fig.~\ref{demarcofig3} for a gas consisting of roughly
$3\times10^4$ atoms at six different temperatures. The Metropolis
algorithm is used to prepare the initial distribution of atoms
among lattice sites at each finite temperature. An additional
harmonic potential characterized by $\alpha=h\times113$~Hz$/\mu
m^2$ is applied to generate a Mott-insulator state with a core of
$n=3$ sites at zero temperature.  This same curvature
$\gamma=113\times2\pi$~Hz$/\mu m^2$ is used for the
state-dependent optical potential while the oscillating magnetic
field is applied. The Schr\"{o}dinger equation is numerically
solved on each lattice site using $\Omega=100\times2\pi$~Hz and
$\omega_{rf}(t)=6.4\times10^4\times2\pi\cdot t$~Hz.  We would like
to have spectral discrimination between sites at different radii,
which limits the Rabi frequency $\Omega$ since it sets the
effective bandwidth for transfer between states.  Therefore we
chose $\Omega$ so that it is comparable to the difference in
$\omega_{ab}$ between sites spaced radially by $a$ near the
minimum of the optical potential. The sweep rate for $\Omega$ is
chosen so that Landau-Zener transitions are made with high
probability. The data in Fig.~\ref{demarcofig3} would be generated
using a frequency sweep lasting $\sim1.26$ seconds, which should
be comparable with spontaneous scattering times and much faster
than loss caused by collisions with residual background gas
molecules.

From the simulation results shown in Fig.~\ref{demarcofig3}, it is
clear that the effects of a state-dependent interaction persist to
relatively high temperature compared with the interaction energy.
The smallest frequency at which atoms are transferred to
$|b\rangle$ decreases with increasing temperature as the size of
the gas in the harmonic potential grows.  In principle, the
temperature of the gas can be inferred from a fit of data obtained
using this technique to the simulation results.

\section{Concluding remarks}
\label{sec:conclusion}

The possibility of exploring novel quantum phases of matter using
ultra-cold atoms trapped in optical lattices has piqued the
interest of many researchers. In this paper, we have detailed the
necessary conditions for creating and probing a shell structure of
Mott-insulator phases in an optical lattice. Instability of the
Mott-insulator shell structure arises principally from two
effects: finite tunneling and finite temperature. Finite tunneling
of the bosons between neighboring sites leads to superfluid
regions between the shells and particle-hole defects in the
Mott-insulator regions. We have shown that for reasonable
experimental parameters, these effects can be made negligible. The
relevant conditions are both that the tunneling parameter must be
small compared with the on-site repulsion ($w/U \ll 1$) and small
compared with the typical change in the confining potential $V$
between neighboring sites ($w/\Delta V \ll 1$) across the
Mott-insulator boundary. At finite temperature, while the Mott
state formally ceases to exist, we have found a reasonable
suppression of defects for temperatures lower than $U/(10 k_b)$
from Monte Carlo simulations of $3 \times 10^4$ atoms. The
existing literature on quantum tunneling at finite temperature
provides a starting point for a more detailed exploration of these
issues in this rich system, for instance, the effects of the
coupling between a single site defect and the rest of the system.
In this paper we have also proposed an experimental probe that has
the capacity to spatially resolve the Mott-insulator shell
structure.

B.~DeMarco acknowledges support from the U.S. Office of Naval
Research, award N000140410490. C.~Lannert and S.~Vishveshwara
thank the KITP, where this work was initiated, for its hospitality
and for valuable discussions.  The authors would also like to
thank A.J.~Leggett for fruitful discussions.


\begin{thebibliography}{99}

\bibitem{Jaksch98} D. Jaksch, C. Bruder, J.I. Cirac, C.W. Gardiner and P. Zoller, Phys. Rev. Lett. {\bf 81}, 3108
(1998).

\bibitem{Kuklov04a} A. Kuklov, N. Prokof'ev, and B. Svistunov,
Phys. Rev. Lett. {\bf 92}, 050402 (2004).

\bibitem{Jane03} E. Jane, G. Vidal, W. Dur, P. Zoller and J.I. Cirac, Quant. Inf. Comp. {\bf 3}, 15
(2003).

\bibitem{Oosten01} D. van Oosten, P. vn der Straten, and H.T.C.
Stoof, Phys. Rev. A {\bf 63}, 053601 (2001).

\bibitem{Kashurnikov02} V.A. Kashurnikov, N.V. Prokof'ev, and B.V.
Svistunov, Phys. Rev. A {\bf 66}, 031601(R) (2002).

\bibitem{Batrouni02} G.G. Batrouni, V. Rousseau, R.T. Scalettar, M. Rigol, A.
Muramatsu, P.J.H. Denteneer, and M. Troyer, Phys. Rev. Lett. {\bf
89}, 117203 (2002).

\bibitem{Wessel04}
S. Wessel, F. Alet, S. Trebst, D. Leumann, M. Troyer, and G.
Batrouni, cond-mat/0411473 (2004).

\bibitem{Rabl03} P. Rabl, A.J. Daley, P.O. Fedichev, J.I. Cirac and P. Zoller, Phys. Rev. Lett. {\bf 91}, 110403 (2003).

\bibitem{Brennen03a} G.K. Brennen, G. Pupillo, A.M. Rey, C.W. Clark and C.J. Williams, quant-ph/0312069 (2003).

\bibitem{LeggettRMP}
A. J. Leggett, S. Chakravarty, A. T. Dorsey, M. P. A. Fisher, A.
Garg, and W. Zwerger Rev. Mod. Phys. {\bf 59}, 1 (1987).

\bibitem{Sethna}
J. P. Sethna, Phys. Rev. B {\bf 24}, 698 (1981); J. P. Sethna,
{\it ibid.} {\bf 25}, 5050 (1982); C. P. Flynn and A. M. Stoneham,
{\it ibid.}, {\bf 1}, 3966, (1970); H. Teichler and A. Seeger,
Phys. Lett. A {\bf 82}, 91 (1981).

\bibitem{HanggiTalknerBorkovec90}
P.~H\"anggi, P.~Talkner, and M.~Borkovec, Rev. Mod. Phys. {\bf
62}, 251 (1990).

\bibitem{Rom04} T. Rom, T. Best, O. Mandel, A. Widera, Ma. Greiner,
T.W. H\"{a}nsch, and I. Bloch, Phys. Rev. Lett. {\bf 93}, 073002
(2004).

\bibitem{Vala03} J. Vala, A.V. Thapliyal, S. Myrgren, U. Vazirani, D.S. Weiss and K.B. Whaley, quant-ph/0307085 (2003).

\bibitem{Greiner02a} M. Greiner, O. Mandel, T. Esslinger, T.W. Hansch and I. Bloch, Nature {\bf 415}, 39 (2002).

\bibitem{FisherWeichmanGrinsteinFisher89}
M. P. A. Fisher, P. B. Weichman, G. Grinstein, and D. S. Fisher,
Phys. Rev. B {\bf 40}, 546 (1989).

\bibitem{GreinerThesis}
M. Greiner, Ph.D. Dissertation, Ludwig-Maximilians-Universit\"at
at M\"unchen (2003).

\bibitem{Sachdev}
S. Sachdev,
{\it Quantum phase transitions\/}
(Cambridge University Press, 1999).

\bibitem{NielsenChuang00}
M. Nielsen and I. Chuang, {\sl Quantum Computation and Quantum
Information\/} (Cambridge Univ. Press, 2000).

\bibitem{dickersheid03} D.B.M. Dickerscheid, D.van Oosten, P.J.H.
Denteneer, and H.T.C. Stoof, Phys. Rev. A {\bf 68}, 043623 (2003).

\bibitem{Meystre2004} C.P. Search, W. Zhang, and P. Meystre, Phys. Rev. Lett. {\bf 92}, 140401 (2004).

\bibitem{BenKish2003} A. Ben-Kish, B. DeMarco, V. Meyer, M. Rowe, J. Britton, W.M. Itano, B.M. Jelenkovic, C. Langer, D. Leibfried, T. Rosenband and D.J. Wineland, Phys. Rev. Lett. {\bf 9}0, 037902 (2003).

\bibitem{Leibfried97a} D. Leibfried, D.M. Meekhof, C. Monroe, B.E. King, W.M. Itano and D.J. Wineland, J. Mod. Opt. {\bf 44}, 2485 (1997).

\bibitem{Meekhof96a} D.M. Meekhof, C. Monroe, B.E. King, W.M. Itano and D.J. Wineland, Phys. Rev. Lett. {\bf 76}, 1796 (1996).

\bibitem{Hall98a} D.S. Hall, M.R. Matthews, C.E. Wieman and E.A. Cornell, Phys. Rev. Lett. {\bf 81}, 4532 (1998).

\bibitem{Volz03}
T. Volz, S. Durr, S. Ernst, A. Marte and G. Rempe,  Phys. Rev. A
{\bf 68}, 010702(R) (2003).

\bibitem{Marte02} Marte, T. Volz, J. Schuster, S. Durr, G. Rempe, E.G.M. van Kempen and B.J. Verhaar,  Phys. Rev. Lett. {\bf 89}, 283202 (2002).

\bibitem{Curtis02} J.E. Curtis, B.A. Koss and D.G. Grier, Opt. Comm. {\bf 207}, 169 (2002).

\bibitem{Sinclair04} G. Sinclair, P. Jordan, J. Leach, M.J.
Padgett, and J. Cooper, J. Mod. Opt. {\bf 51}, 409 (2004).



%
%\bibitem{CaldeiraLeggett81}
%A. O. Caldeira and A. J. Leggett, Phys. Rev. Lett. {\bf 46}, 211
%(1981), and references therein.
%
%\bibitem{Coleman}
%Coleman; Callen and Coleman; Phys. Rev. D
%
%\bibitem{Kramers}
%H. A. Kramers, Physica {\bf 7}, 284 (1940).

\end{thebibliography}
\end{document}